\documentclass[conference]{IEEEtran}
\usepackage{amsmath,amssymb,amsthm,epsfig,color,subfigure,empheq,graphicx,graphics,pgfplots}

\usepackage{enumerate,url,algorithm,algorithmic,wasysym,epstopdf,enumitem}
\usepackage{siunitx}
\usepackage{graphicx}
\usepackage{multicol}
\usepackage{makecell}
\graphicspath{ {./template/} }
\usepackage{tikz}
\usepackage{comment}
\def\BibTeX{{\rm B\kern-.05em{\sc i\kern-.025em b}\kern-.08em
    T\kern-.1667em\lower.7ex\hbox{E}\kern-.125emX}}
\usetikzlibrary{shapes.geometric, arrows}
\tikzstyle{startstop} = [rectangle, rounded corners, minimum width=3cm, minimum height=1cm,text centered, draw=black, fill=red!30]
\tikzstyle{io} = [trapezium, trapezium left angle=70, trapezium right angle=110, minimum width=3cm, minimum height=1cm, text centered, draw=black, fill=blue!30]
\tikzstyle{process} = [rectangle, minimum width=3cm, minimum height=1cm, text centered, draw=black, fill=orange!30]
\tikzstyle{decision} = [diamond, minimum width=3cm, minimum height=1cm, text centered, draw=black, fill=green!30]
\tikzstyle{arrow} = [thick,->,>=stealth]
\begin{document}
\title{Undetectable Relentless Incremental GPS-Spoofing Attack on Time Series PMU Measurements and Detection Model}

\author{\IEEEauthorblockN{Imtiaj Khan}
\IEEEauthorblockA{\textit{Electrical and Computer Engineering} \\
\textit{Virginia Tech}\\
Blacksburg, VA, USA \\
imtiajkhan@vt.edu}\\
\and
\IEEEauthorblockN{Virgilio Centeno}
\IEEEauthorblockA{\textit{Electrical and Computer Engineering} \\
\textit{Virginia Tech}\\
}

}
\maketitle

\begin{abstract}
The Phasor Measurement Unit (PMU) is an important metering device for smart grid. Like any other Intelligent Electronic Device (IED), PMUs are prone to various types of cyberattacks. However, one form of attack is unique to the PMU, the GPS-spoofing attack, where the time and /or the one second pulse (1 PPS) that enables time synchronization are modified and the measurements are computed using the modified time reference. This article exploits the vulnerability of PMUs in their GPS time synchronization signal. At first, the paper proposes an undetectable gradual GPS-spoofing attack with small incremental angle deviation over time. The angle deviation changes power flow calculation through the branches of the grids, without alerting the System Operator (SO) during off-peak hour. The attacker keeps instigating slow incremental variation in power flow calculation caused by GPS-spoofing relentlessly over a long period of time, with a goal of causing the power flow calculation breach the MVA limit of the branch at peak-hour. The attack is applied by solving a convex optimization criterion at regular time interval, so that after a specific time period the attack vector incurs a significant change in the angle measurements transmitted by the PMU. Secondly, while the attack modifies the angle measurements with GPS-spoofing attack, it ensures the undetectibility of phase angle variation by keeping the attack vector less than attack detection threshold. The proposed attack model is tested with Weighted Least Squared Error (WLSE), Kalman Filtering, and Hankel-matrix based GPS-spoofing attack detection models. Finally, we have proposed a gradient of low-rank approximation of Hankel-matrix based detection method to detect such relentless small incremental GPS-spoofing attack. The results demonstrate the failure of WLSE and Kalman Filtering models in the detection of proposed relentless GPS-spoofing attack. While Hankel-matrix based model provides insights on phase angle variation during such attack, the proposed gradient of low-rank approximation of Hankel-matrix based detection model can perform significantly better and can detect the proposed attack model conclusively. We tested the proposed attack schemes and the detection models with simulated measurements of IEEE 24 bus RTS om PSS/E and MATLAB.
\end{abstract}


% \fi
\IEEEpeerreviewmaketitle

\section{Introduction}

Introducing smart metering and protective devices brought a revolution in the traditional power grid. A smart grid comprises of two parts: a physical part which is the conventional power grid and a control center which receive data and takes decision based on the data it received, combination of both part is referred as Cyber-Physical System (CPS). In a typical CPS, Supervisory Control And Data Acquisition (SCADA) and Advanced Metering Infrastructure (AMI) are utilized in a two-way communication system between physical power system and  control center at the cyber layer \cite{i1}. The control center monitors any fluctuation or  economic dispatch, unit commitment and optimal power flow (OPF) \cite{i2}.

More advanced and complicated CPS require fast and accurate Intelligent Electronic Devices (IEDs), one such example is Phasor Measurement Unit (PMU), which is relatively newer type of metering device providing voltage and current magnitudes and phase angles in a time synchronized format \cite{i3}. The measurement data provided by PMUs are time-tagged with Universal Time Coordinate (UTC) and the time information include second-of-century (SOC) count, fraction-of-second (FRACSEC) count, and time quality flag. The time information is synchronized with a reliable time-source and the most common time-source is the Global Positioning System (GPS). In addition to the magnitudes and phase angles of voltage and current components, PMUs also provide frequency and Rate of Change of Frequency (ROCOF) data \cite{i4}. Installing PMU in the smart grid created a new horizon in terms of accuracy and speed \cite{i5} \cite{i6}. In conventional SCADA based system, data transfer rate is one sample per second, where in PMU the data transfer rate is 30 to 120 sample per second \cite{i7} \cite{i8}.

Despite being reliable in terms of time-tagged data and accuracy, PMUs are still susceptible to various type of cyber attacks \cite{i9} \cite{i10} \cite{i11}. The most common type of cyber attack against PMU, similar to other IEDs and network system, is the malicious data injection attack such as false data injection attack (FDIA). Other common types of cyberattacks include data modification and replay attacks \cite{i12}. In FDIA, the attacker injects bad data into the system and thereby modifying the power system data fed into the control centre. If the state estimation is performed with falsified data, the System Operator (SO) at control centre may take wrong decision such as taking restorative action or emergency shut down even though the operating condition is not actually violated. Another frequent type of attack that can hamper PMU integrated grid performance is Denial of Service (DoS) attack. In this case the attacker injects large volume of data through the communication channel, therefore the data transfer to destination gets blocked due to the large volume of data traffic \cite{i13} \cite{i24}. Man In the Middle (MITM) and Side Channel attacks are also critical for PMU \cite{i14} \cite{i15}.

Since PMU relies on GPS signal for time synchronization, it opens the possibility of a new type of attack: GPS-spoofing attack. PMU generally uses public GPS which lacks the sophisticated protection scheme used for military GPS. GPS-spoofing attack poses very serious concern over the cybersecurity of the CPS \cite{i17}. It may affect the magnitude and the phase angle, however the phase angle is the most susceptible portion of the synchrophasor, because shift in the GPS 1 PPS, which is used as the common reference for all PMUS, is reflected by the phase angle shift \cite{i16}. Slowly changing time reference for an individual PMU slowly shift the phase angle data \cite{i18} \cite{i19} provided by the particular PMU from a particular node. Gradual change of phase angle of a particular node with respect to the other nodes in the grid can, in some scenarios, provide incorrect information to a system operator (SO) at control centre that may lead the SO take unnecessary actions. Moreover, the gradual shift in phase angle measurements due to the GPS-spoofing attack has the potential to impact any PMU base transmission line fault detection and identification of event location \cite{i17}.

The phase angle shift caused by the GPS-spoofing attack can affect a PMU based transmission line differential protection scheme. As soon as the spoofed phase angle crosses the threshold, the relay will trip despite the absence of an actual violation of the operating constraint. In this way the attacker can disrupt the normal operating condition of the grid, without creating any physical change in the cyber-physical model. This phenomenon can be referred as misoperation \cite{i25}.

Several researchers focused on the protection of PMU integrated smart grid against cyber attacks. Most works are based on detecting FDIA. Detection of FDIA is similar to conventional Bad Data Detection (BDD) method \cite{i20} \cite{i21}. Conventional BDD algorithms observe the residuals of the measured and expected variables and do statistical test to find the outliers. GPS-spoofing attack can also be considered as a type of BDD, since modification in time reference lead to the shift in phase angles. In GPS-spoofing attack, the voltage and current magnitude may remain unchanged, depending on the algorithm employed by the PMU manufacturer. As a result, it can be considered similar to a FDIA with only the phase angle data modified.

Considering the defense strategy taken by control center, it is possible to create attacks that are undetectable by BDD algorithms \cite{i22} \cite{i23}. This statement is also true for GPS-spoofing attack if the attack is considered as a variation of FDIA with corrupted phase angle data \cite{11}. These types of undetectable and stealthy attacks can still be prevented by placing PMUs into optimal locations of the grid \cite{7}. The goal of this article is to create an GPS-spoofing attack, which cannot be prevented by optimal PMU placement or by BDD algorithms. The key idea is to inject a shift in the phase angle by creating and maintaining an effective delay in time reference of an individual PMU. Doing so, the attacker will be able to change the phase angle difference between the spoofed PMU and any other monitored PMU in the system. The sample attack used in this article is made in such a way that, its impact on the power flow calculation is insignificant for one instance of attack. The attack is incrementally applied after some interval of time for over a specific time period T.

The attackers’ goal is to eventually cause a significant impact on the SO’s perceived power flow measurement after the time period T. During off-peak hour, the power flow in the line under attack is well below its limit, as indicated by the angle difference of the PMUs at the both ends of the line. The SO will not suspect the power flow to exceed the limit at this time, therefore a significant increase in the calculated power flow will alert the SO about possible cyberattack. The attacker exploits this issue and over the period T, at each time instance s/he creates insignificant increase in the phase angle of a PMU at one end of the line, causing an insignificant increase in the derived power flow through the line. The accumulative effect of the increases grows larger over time, and after T, which can be made to coincide with the peak load, if performed correctly the line power flow as determined by the phase angle difference from PMU measurements exceeds the line’s limit. At peak load-hour, the SO at the control centre is prepared for possible operational limit violations in the physical grid, therefore if the calculated power flow exceeds the line low limit at that time, the SO, considering a physical cause for the event, may then take the protocols required for operational limit violation despite the operational limit is not actually violated. Therefore, the load-shedding or other restorative measures will be taken, which will lead to a hamper in the supply of power to the critical points of the grid.

The attacker starts the attack at the initial time $t_0$. The phase angle shift $a_0$ generated by the spoofed signal will be derived by solving optimization equation which aims at enabling the attack undetectable by commonly used detection techniques that are discussed in section III. The attacker creates a spoofed GPS signal to the PMU at bus $i$ and/ or $j$ at $t_0$, introducing a time shift $\Delta t_0=\frac{a_0}{2 \pi f}$ \cite{r10}, corresponding to the optimal phase angle measurement shift $a_0$  caused by spoofing and system frequency $f$. The fast data transmission rate of PMUs, along with the time required by optimization equation for undetectable attack, put a constrain on the attacker regarding the knowledge of the target angle $\theta_i$ and $\theta_j$. Assuming the time required to execute optimization equation is $T_s$ sec., the attacker is required to have the knowledge of $\theta_{i0}$ and $\theta_{j0}$ at least $T_s$ sec. ago. It is practically impossible for the attacker to have the exact phase angle value of a future timestamp. To circumvent this issue, attacker can exploits data estimation or recovery techniques to accurately predict the phase angles of the timestamp which is $T_s$ sec. later. At the timestamp $t_0 - T_s$, considering the $T_s$ sec. of computational time-lag, the proposed attack model estimates the phase angles $\theta_{i0}$ and $\theta_{j0}$ of timestamp $t_0$. Attacker solves the optimization equation for undetectable attack to generate the optimal phase angle shift $a_0$ using the estimated $\theta_{i0}$ and $\theta_{j0}$.

%%%%%%%%%%%%%%%%%%%%%%%%%%%%%%%%%%%%%%%%%%%%%%%%%%%%%%%%%%%%%%%%%%%%%%%%%%%%%%%%
All the sampled measurements of PMU following the timestamp $t_0$ will carry on this time shift in their measurements and the corresponding phase angle measurements will be phase shifted by $\Delta \theta_0  = 2 \pi f \Delta t_0 = a_0$. The next attack is initiated at time $t_1$, after S number of samples. At time $t_1 - T_s$, the attacker estimates the phase angle measurements $\theta_{i1}$ and $\theta_{j1}$ for timestamp $t_1$. At $t_1$, the attacker solves the optimization equation using previously estimated phase angle measurements, ($\theta_{i1}$ and $\theta_{j1}$), to compute the new attack value $a_1$. The corresponding time shift that the attacker needs to introduce by GPS-spoofing will be $\Delta t_1  = \frac{a_1}{2\pi f}$ . The overall phase angle shift in the PMU measurement for the next S number of samples becomes $\Delta \theta_1  = 2 \pi f \Delta t_1  + \Delta \theta_0$. The power flow through the branch $i - j$ will also deviate accordingly. The attacker keeps instigating the spoofing attack for the following timestamps $t_2,t_3,.... \text{until  T}$, at which the power flow through the branch $i - j$ reaches the peak value. The tempered power flow, which kept increasing from the actual power flow in the previous timestamps, suffers from maximum deviation and crosses the power flow limit of the line $i - j$.

\begin{figure}[hbt!]
 \includegraphics[width=0.5\textwidth]{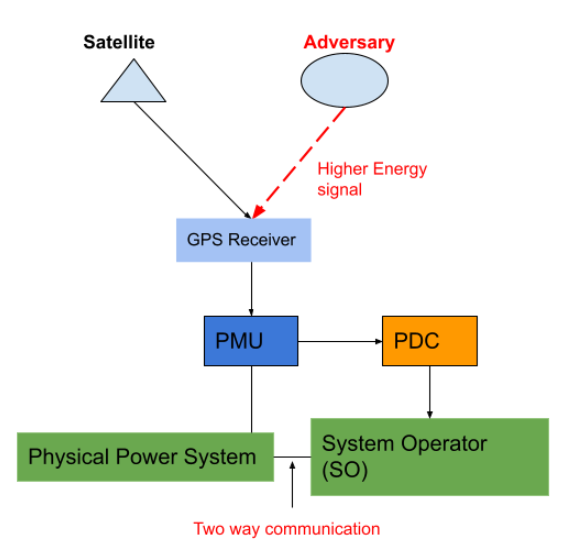}
 \caption{GPS spoofing attack structure}
 \label{spoof}
\end{figure}

The main difference between the GPS-spoofing attack of the proposed model and FDIA is that, in the conventional FDIA the attacker needs to get into the cyber layer or breach network security to initiate the attack. In this model the attacker is not required to breach network security, the attack is launched by spoofing the GPS signal received by the PMU, and the attack is outside of the cyber-physical system of the smart grid. In FDIA, the attacker injects the false data into the cyber layer; however, in the GPS spoofing attack model discussed in this work, the attacker doesn’t inject any falsified measurement, rather s/he forces the PMU algorithm to introduce a phase shift in the computed phasor by disrupting GPS signal received by the PMU’s GPS receiver. Instead of altering the phase angle measurement directly, the GPS spoofing attack shifts the 1 Pulse-Per-Second (PPS) signal used by the PMU for time reference, and this shift in 1 PPS reference signal is reflected by a small change in phase angles for the phasors computing until the next 1 PPS is received. The overall topological structure of the GPS-spoofing attack considered in this work can be depicted in fig \ref{spoof}.

The contribution of this work has four aspects: firstly, we utilize a Hankel-matrix based data recovery technique, that estimates phasor measurements of the future timestamp which is $T_s$ later. Secondly, with the pre-estimated phasor measurements, the attacker calculates the phase angle shifts using optimization equation for each 1 PPS of the GPS receiver with a goal of keeping the small individual change in phase angle measurement undetectable by most bad data detection algorithms. The constraints of this optimization method is similar to stealthy FDIA methods discussed in previous literature. However, existing stealthy FDIA algorithms considered the attack vector to be greater than a significance threshold $\zeta$, whereas the proposed method in this article consider the attack vector to be less than $\zeta$ to create insignificant variation in the phase angle measurements at each time instance. Our proposed model creates an attack that has the objective of slowly shifting a PMU angle to achieve a maximum impact at the peak loading time of a line monitored by two PMUs. In conventional FDIA, attack values at each timestamp are discrete from the previous ones; however, in the proposed GPS-spoofing model the phase shift $\theta_m$ at each affected timestamp $t_m$ is maintained over 1 PPS and added up to the phase shifts incurred by the previous attacks $\sum_{k=1}^m \theta_{k-1}$  that was introduced at the corresponding 1 PPS signals at the timestamps $t_0,t_1,...t_{m-1}$. Finally, we have proposed a novel gradient of low-rank approximation error of Hankel-matrix based relentless incremental GPS-spoofing attack detection method and verified the effectiveness of proposed detection model.

The paper is organized as follows: Section II briefly discusses the data estimation technique utilizing Hankel-matrix based missing data recovery model. Section III explains the proposed undetectable attack scheme.% Section IV validates the undetectibility criteria of the proposed method.
The conventional Bad Data Detection (BDD) techniques as well as proposed novel gradient of low-rank approximation error of Hankel-matrix based detection model is discussed in section IV. The test setup for the proposed stealthy attack is discussed in section V. Section VI provides the results depicting the impact of the proposed GPS-spoofing attack on the power flow calculations as well as analyzes the performances of several BDD models, including the novel gradient based model. Section VI concludes the work.

The main contribution of this article can be summarized as follows:
\begin{itemize}
    \item A relentless small incremental GPS-spoofing attack model is proposed targeting the power flow calculations;
    \item An optimization criterion is proposed to create an attack that is undetectable against conventional BDD. and can incur a significant impact on the perceived power flow between two nodes during peak load-hour by assuming an angle stability limit in the constraint, even though the actual flow through the same branch remain unchanged;
    \item The undetectibility of the proposed GPS-spoofing attack at each affected timestamp is tested against multiple BDD algorithms such as WLSE, Kalman Filtering, and Hankel-matrix. Measurements with noise from the IEEE 24 bus RTS in PSS/E are used to generator phasor measurements;
    \item A gradient of low-rank approximation error of Hankel-matrix based model is proposed and tested to detect such relentless small incremental GPS-spoofing attack model;
    
\end{itemize}

\section{Estimation of Future Measurements}\label{miss_rec}

The first step of the proposed model is the estimation of phasor measurements for future timestamp. Pre-estimation of measurements for future timestamp works similarly as missing data recovery techniques. Several missing data recovery models can be found in literature, such as spatial and temporal OnLine Algorithm for Phasor data recovery models \cite{rec1} \cite{r29}, singular value thresholding \cite{rec3}, JPT algorithm \cite{rec4}, supervised cascaded convolutional auto-encoders \cite{rec5}, Iterative Adaptive Approach (IAA) \cite{rec6}, tensor decomposition \cite{rec7} etc. Even though the methods proposed in \cite{rec3} - \cite{rec7} provide accurate missing data recovery, they are applicable for off-line recovery only. Online algorithm with temporal singular vector proposed in ref \cite{r29} provide relatively higher accuracy than the algorithm with spatial singular vector \cite{rec1}. A more advantageous missing data recovery method was proposed in \cite{rec8} using low-rank Hankel structure. This method has the ability to estimate larger number of missing data than the method proposed in \cite{r29}. Furthermore, the Hankel-matrix based algorithm can correct consecutive bad data as well as help the system operator distinguish GPS-spoofing attack from conventional FDIA \cite{rec9}.
The Hankel-matrix based missing data recovery model takes phasor measurements from PMUs as input, and subsequently perform low rank approximation to estimate the next measurement. The first step of Hankel-matrix based model is to create measurement matrix \textbf{Z}, containing phasor measurements over $\tau$ consecutive time instances as in eqn. \ref{e1}.

\begin{align}\label{e1}
\begin{matrix}
\textbf{Z} = [z_1   z_2 ....   z_\tau]
\end{matrix}
\end{align}

The Hankel-matrix is created by shifting each row to one data point left from the previous row. From the measurement matrix $\textbf{Z}$, the Hankel-matrix $H$ can be created with varying number of rows $\kappa$, constituting a $\kappa \times \tau-\kappa+1$ matix as in eqn \ref{hankel}.

\begin{align}\label{hankel}
H = \begin{bmatrix}
z_1 & z_2 & ... & z_{\tau-\kappa+1}\\
z_2 & z_3 & ... & z_{\tau-\kappa+2}\\
... & ... & ... & ...\\
... & ... & ... & ...\\
z_\kappa & z_{\kappa+1} & ... & z_\tau
\end{bmatrix}
\end{align}

The low rank approximation is defined as converting the original matrix $\textbf{H}$ into a new matrix with lower rank $r$, where $r < rank(\textbf{H})$. The low rank approximation is performed by taking first r largest singular values from the Singular Value Decomposition (SVD) of \textbf{H}, which is $U \Sigma V^*$. The low rank approximated matrix $\bar{\textbf{H}}$ is created with rank-$r$ matrix $U \Sigma^r V^* $.
The low rank approximation error is defined as eqn \ref{lra}. 

\begin{align}\label{lra}
    e^r(\textbf{H}) = \frac{|| U \Sigma^r V^* - \textbf{H} ||_F}{\textbf{H}}
\end{align}

 To estimate missing data at the time instance $\tau+1$ using the measurements from previous $\tau$ time instances, the state variable $d$ is predicted using the following relation \cite{r29}:

\begin{align}
    \hat{d} = (\hat{U^r}^T \ast \hat{U^r})^{-1} \hat{U^r} \ast \bar{\textbf{H}}
\end{align}

where $\hat{U^r}$ is the first $r$ dominant left singular matrix from $U$. For each PMU channel $i$, where $i = 1,2,...,M$, $M$ being the number of PMU channels, the data at time instance $\tau + 1 $ is estimated from $\hat{U^r} \hat{d}$.

Assuming the PMU channel I contains missing data at the time instance $\tau+1$. The system operator (SO) at the control center fills the missing data at that time instance with estimated measurement. The measurement at time $\tau+1$ is expressed as $\hat{z_{\tau+1}} = \hat{U^r} \hat{d}$.

During physical event or certain cyberattack causing data-drop through communication channel, it is possible to have consecutive missing data in the measurement stram sent by PMUs to the PDCs. To tackle this scenerio, the consecutive missing data is estimated using the previously estimated missing data. To estimate missing data at the time instance $\tau+2$, measurement matrix similar to eqn \ref{e1} is created with measurements stream from $z_2$ to $\bar{z_{\tau+1}}$. The estimated measurement $\bar{z_{\tau+2}}$ is derived using the term $\hat{U^r} \hat{d}$, there $\hat{d}$ is new estimated state variable with updated measurement matrix.

The estimated measurements using low rank approximation of Hankel-matrix provide accurate recovery of consecutive missing data, however, the accuracy declines for very large number of consecutive missing data. A properly tuned Hankel-matrix provides upto $99.8 \%$ accuracy when consecutive missing data constitutes $40\%$ of the measurement matrix \cite{rec8}.

The proposed undetectable GPS-spoofing attack model relies on phasor measurements in real-time to solve the optimization algorithm aiming at shifting the time reference for PMUs. The computational time of the optimization algorithm puts an obstacle to the attacker in the way of spoofing the GPS signal in real-time. Therefore, instead of solving the optimization algorithm in real-time, the attacker take the computational delay into consideration and solve the optimization criterion few timestamps ahead of the actual timestamp when the shift in time reference due to GPS-spoofing attack takes place. It is physically impossible to have the exact measurement information beforehand; as a result, the attacker of the proposed model depends on predicting measurements few timestamps in future and solve the optimization algorithm to figure out the required shift in timestamp to instigate undetectable GPS-spoofing attack.

\section{Formulation of Undetectable Timing Attack}\label{formu}

During GPS-spoofing attack, the time synchronization with GPS signal is distorted and the PMUs start sending measurements of shifted timestamp to PDCs. Due to the change in horizontal time-axis, the most affected part of phasor measurement is the phase angle under GPS-spoofing attack; in contrast to only a slight change in the magnitude measurements. The SO needs to apply bad data detection algorithm on phase angle measurements particularly, rather than looking at the magnitudes, in order to identify the existence of any consecutive bad data caused by GPS-spoofing attack. This requirement can be exploited by the attacker. The key idea is that the attacker needs to shift the time reference by a very small amount, which causes very small deviation in phase angle measurements, subsequently making the attack undetected by the SO. If the attacker keeps shifting the time reference by this small amount after each second over a period of time, there will be a significant amount of shift in time reference after the whole period. From the perspective of SO, it is not possible to detect this shift in GPS 1 PPS time reference by performing state estimation based bad data detection on phasor measurements after each millisecond or fractions of second, since the shift in GPS 1 PPS time reference satisfies the undetectibility constraints of state estimation based method. 

\subsection{Optimizing Attack Vector}

The most common bad data detection method employed at the control center is Weighted Least Squared Error (WLSE) based state estimation. The undetectable attack model in this work focuses on circumventing the WLS based state estimation technique.

For an n bus system, the state variable at a specific time instance can be represented as the vector $\textbf{x} = [x_1, x_2,....x_n]$ and the measurement variables can be represented with the vector $\textbf{z} = [z_1, z_2,....z_m]$. Here m is the number of meters (m$>>$n). The relation between $\textbf{z}$ and $\textbf{x}$ can be expressed as 

\begin{equation}\label{e1}
    \textbf{z} = \textbf{H} \textbf{x} + e
 \end{equation}
 
 H is the Jacobian matrix that represents the non-linear relation between the state variable and the measurement variable. $e = [e_1, e_2,....e_m]$ is the measurement error vector. An FDIA will go undetected if $|| z - H \hat{x} ||_2$ is less than threshold $\tau^r$, $\hat{x}$ being the estimated state variable which can be calculated from the Weighted Least Square (WLS) method $\hat{x} = (H^T R^{-1} H)^{-1} H^T R^{-1} z$. R is the measurement error covariance matrix. If the attack $\textbf{a}$ vector is applied during FDIA, the measurement vector under attack will be $z_a = z + a$ and the estimated state variables under attack will be $\hat{x_a} = \hat{x} +c$. The estimated attack vector is $c = (H^T H)^{-1} H^T a$. After injecting the attack vector, $|| z_a - H \hat{x_a} ||_2$ = $|| z + a - H \hat{x} - H c ||_2$ $\leq$ $|| z - H \hat{x} ||_2$ + $|| a - H c ||_2$. For a carefully crafted attack vector, such as $a = Hc$, the estimation residual becomes $|| z_a - H \hat{x_a} ||_2 \leq \tau^r$. As the estimation residual becomes less than threshold $\tau^r$, the SO fails detect FDIA \cite{7}.

The attacker goal is to make the term $||a - Hc||_2$ as small as possible so that $|| z_a - H \hat{x}_a||_2$ remains less than the threshold $\tau^r$. $a - Hc$ can be written as $F a$, where the term $F = (H(H^T H)^{-1} H^T - I)$. The attacker needs to satisfy the following criterion:

\begin{equation}\label{op2}
\begin{array}{rrclcl}
\displaystyle \min_{a} & ||F a||_2\\
\end{array}
\end{equation}
To avoid getting detected, the term $||F a||_2$ needs to be between $\tau^r$ and $||z-H\hat{x}||_2$ and the attacker must satisfy the following optimizer:

\begin{equation}\label{op20}
\begin{array}{rrclcl}
\displaystyle \min_{a} & ||F a||_2\\
\textrm{s.t.} & ||F a ||_2 \leq \tau^r - ||z-H \hat{x}||_2\\
\end{array}
\end{equation}

Assume the impact threshold of attack vector is $\zeta$, implying if the attack vector is greater than this value, there will be a significant impact on the system, else the attacker cannot inflict any damage to the grid. To avoid detection, the attack vector \textbf{a} must be less than impact threshold during each time instance, however the total impact, after the phase angle deviation at every time instance being added up, will be significant at the end of total time period \textbf{T}. For each time instance, the attacker has to satisfy the following:

\begin{equation}\label{op21}
\begin{array}{rrclcl}
\displaystyle \min_{a} & ||F a||_2\\
\textrm{s.t.} & ||F a ||_2 \leq \tau^r - ||z-H \hat{x}||_2\\
& ||a||_1 < \zeta\\
\end{array}
\end{equation}

Each time instance can be depicted as index i; i = 1,2,....,T. T = total time period. Therefore for the whole time period the optimization criteria becomes:
\begin{equation}\label{op3}
\begin{array}{rrclcl}
i = 1:T & \{\\
\displaystyle \min_{a_i} & ||F a_{:,i}||_2\\
\textrm{s.t.} & ||F a_{:,i} ||_2 \leq \tau^r - ||z_{:,i}-H \hat{x}_{:,i}||_2\\
& ||a_{:,i}||_1 < \zeta\\
\}\\
\\
& \sum_{i=1}^{T} ||a_i||_1 > \zeta\\
\end{array}
\end{equation}

The next challenge is to find the impact threshold $\zeta$. The $\zeta$ must be less than the $\ell_1$ norm of the attack vector for single time instance to avoid detection, but it must be greater than the sum of the $\ell_1$ norm of the attack vector over the total T period of time to incur damage to grid operation. Since the $||a||_1$ can takes the value between 0 to $\zeta$ at each time instance, it should be between $\zeta/T$ and $\zeta$ so that the $\sum_{i=1}^{T} ||a_i||_1 > \zeta$ condition is satisfied.

The optimizer \ref{op3} becomes:
\begin{equation}\label{op4}
\begin{array}{rrclcl}
i = 1:T & \{\\
\displaystyle \min_{a_i} & ||F a_{:,i}||_2\\
\textrm{s.t.} & ||F a_{:,i} ||_2 \leq \tau^r - ||z_{:,i}-H \hat{x}_{:,i}||_2\\
& \zeta/T < ||a_{:,i}||_1 < \zeta\\
\}\\
\end{array}
\end{equation}

In the optimization criteria \ref{op4}, mentioning the suffix $i$ is redundant, since if the attack vector at each time instance satisfies $\zeta/T < ||a_{:,i}||_1 < \zeta$, the impact will be automatically greater than threshold over the whole time period T. Additionally, $\zeta/T$ can be replaced as $\zeta'$. Since the eqn \ref{op4} is a convex, there exists a global minima that corresponds to the attack vector satisfying undetectibility constraint. For the second inequality constraint, the solver needs to search the point in a wide range of values between $\zeta'$ and $\zeta$, depending on the size of time period T. The second inequality constraint can be computationally simplified by making the attack vector within the range $\zeta'$ to $\zeta' + \epsilon$, where $\epsilon$ is a small positive integer. In this case, the the solver needs to search for the optimum attack vector within small range of constraints.

\subsection{Determining Impact Threshold}
The next challenge is to find the impact threshold $\zeta'$. The impact must not exceeds the operating limit of the power system. For an instance, if the operating constraint is violated, the control center will be aware of the situation earlier than the expected and will take restorative action before the target is achieved by the attacker. The operating limits of the power system are:\\
Branch current limit between the bus i and bus j,\\

$-I_{lim} \leq I \leq I_{lim}$\\

Power flow limit between the bus i and bus j,\\

$-P_{lim} \leq P \leq P_{lim}$\\

The power flow between two buses generally depends on the voltage magnitudes of the two nodes, the phase angles of two nodes and the line admittance between the nodes. In AC power flow model, real power flow between the bus i and bus j will be:\\
\begin{equation}
\begin{gathered}
    P_{ij} = V_i^2 (g_{si} + g_{ij}) - V_i V_j [g_{ij} cos(\theta_i - \theta_j) +\\ b_{ij} sin(\theta_i - \theta_j]  
\end{gathered}
\end{equation}

$g_{si}$ is the shunt conductance at bus i, $g_{ij}$ is the line conductance between bus and bus j, and $b_{ij}$ is the line susceptance between bus i and bus j. $V_i$ and $V_j$ are the voltages at bus i and j respectively.
For simplicity, the DC power flow model has been considered in this work. In the DC power flow model the voltage phase difference are very small $(\theta_i \approx \theta_j)$ and the voltage magnitudes at buses i and j are normalized to unit. The state variable x constitutes with the voltage phasors $(x = [ V_1 \angle \theta_1, V_2 \angle \theta_2,....V_n \angle\theta_n])$; n = total number of buses. Only power that comes into consideration is the real power in DC power flow model. The reactance of the line between i and j is $x_{ij}$ and the line resistance is $r_{ij}$, as a result the susceptance $b_{ij} = \frac{-x_{ij}}{r^2_{ij} +x^2_{ij}}$. The line resistance is considered to be negligible in DC power flow model, which implies $r << x$. Hence the susceptance can be approximated as $b_{ij} \approx - 1/x_{ij}$.\\
The power flow between the bus i and j in DC model can be expressed as follows:

\begin{align}
\begin{gathered}
    P_{ij} = -V_i V_i sin(\theta_i - \theta_j) b_{ij}\\
    \approx sin(\theta_i - \theta_j)/x_{ij}\\
    \approx \frac{\theta_i - \theta_j}{x_{ij}}
\end{gathered}
\end{align}

In the attack model described in the previous subsection, attack vector a consists of two elements (since two phase angles are required to be shifted in order to create meaningful attack vector\cite{11}). The $\ell_0$ norm of vector $a$ is 2. Elements of attack vector a represents the shift in voltage phase angle at different buses. Assuming two attack targets bus i and bus j, the corresponding shift in phase angles are represented by $a_i$ and $a_j$. After the attack, the new power flow will be:

\begin{align}
    \begin{gathered}
    P'_{ij} = \frac{\theta_i + a_i - \theta_j - a_j}{x_{ij}}\\
    = \frac{\theta_i - \theta_j}{x_{ij}} + \frac{a_i - a_j}{x_{ij}}\\
    = P_{ij} + \frac{a_i - a_j}{x_{ij}}
    \end{gathered}
\end{align}

To make the power flow between bus i and bus j within the power flow limit, $P'_{ij}$ must be less than $|P_{lim}|$. The second constraint of the eqn \ref{op21} can be rewritten as:

\begin{align}\label{const1}
    \begin{gathered}
    \frac{a_i - a_j}{x_{ij}} < |P_{lim}| - \frac{\theta_i - \theta_j}{x_{ij}}
    \end{gathered}
\end{align}

The right hand term of the eqn \ref{const1} can be approximated as the impact threshold $\zeta'$. The optimization criterion from eqn \ref{op4} is modified to the following form:

\begin{align}\label{final}
\begin{gathered}
\begin{array}{rrclcl}
t = 1:T & \{\\
\displaystyle \min_{a_i} & ||F a_{:,i}||_2\\
\textrm{s.t.} & ||F a_{:,i} ||_2 \leq \tau^r - ||z_{:,i}-H \hat{x}_{:,i}||_2\\
& \zeta' < ||a_{:,i}||_1 < \zeta' + \epsilon\\
&\epsilon > 0\\
\}\\
\end{array}\\
\\
\textrm{where}, \zeta' = x_{ij} \left[|P_{lim}| -\frac{\theta_{i,t} - \theta_{j,t}}{x_{ij}} \right]/T
\end{gathered}
\end{align}

To implement the attack, the attacker needs to solve the eqn. \ref{final} at every time instance $t_i$, that puts an obstacle for the attacker with the computational time required by eqn. \ref{final}. Therefore the attacker needs to predetermine the attack vector, specifically few times instance ahead of $i$, to inflict the GPS-spoofing attack at exactly $i$. $T_s$ being the maximum computational time required to solve eqn. \ref{final}, the attacker needs to determine the attack vector at least \textbf{a} $t_i - T_s$ time instance to implement the corresponding GPS 1 PPS shift at exactly $t_i$. If there exists $K$ number of measurements between $t_i$ and $T_s$, the attacker must predict the following $K$ phase angle measurements accurately to have prior knowledge about the phase angle at $t_i$. To predict the $K$ number of data, the proposed attack model exploits missing data recovery model described in section \ref{miss_rec}.

\section{Test of Undetectibility}

GPS-spoofing attacks impact the phase angle measurements most, therefore it can be considered as FDIA targeting phase angle measurements only \cite{1}. System Operator (SO) looks for the uccessive bad data in the measurements to suspect the occurance of FDIA, which makes FDIA detection similar to Bad Data Detection (BDD). As discussed in section \ref{formu}, the most common Bad Data Detection technique is the WLSE based state estimation method \cite{r26}, where the System Operator (SO) receives the measurements from the physical power system and calculates the residuals between the actual measurement $z$ and the estimated measurement $H \hat{x}$. The notation z, H and x are same as discussed in \ref{formu}. The SO considers the measurements as suspicious if the residual $||r||_2 = ||\hat{z} - H \hat{x}$ is less than a predetermined threshold $\tau^r$.

Kalman Filtering (KF) \cite{r25} is an effective tool to detect bad data caused by electrical events and FDIA. For a system, if the state variable is $x(t)$ and the measurement variable is $y(t)$, the state variable at time $t+1$ can be estimated from the previous state $x(t)$ using the following relation:

\begin{equation}
    x(t+1) = A x(t) + w(t)
\end{equation}

%A is the state transition matrix and w is the system noise. The relation between state variable and measurement is as follows:
\begin{comment}

\begin{equation}
    z(t) = H x(t) + e
\end{equation}

H is defines the relation between the state variables and measurements, similar to eqn \ref{e1} and $e$ is the measurement noise. $w$ and $e$ are zero-mean Gaussian signals and the corresponding covariance matrices are Q and R respectively. 
\end{comment}
KF estimates the state variables $\hat{x}(t)$ at time t using the state variables and measurements upto t-1. P(t) is the covariance of estimates at time t. The time updates of state variables and covariance matrix are expressed as:

\begin{align}
    x(t+1|t) = A x(t) + w(t)\\
    P(t|t-1) = A P(t-1) A^T + Q
\end{align}

Consequently, the measurements are updated using the following relations:

\begin{align}
    K(t) = P(t|t-1) H^T [ H P(t|t-1) H^T +R]^{-1}\\
    P(t|t) = P(t|t-1) - K(t) H P(t|t-1)\\
    \hat{x}(t) = A \hat{x}(t-1) + K(t) [z(t) - H \hat{x}(t|t-1)]\\\label{KFr}
    P(t) = [I - K(t) H] P(t|t-1)
\end{align}

$K(t)$ is the Kalman gain at time t. Starting from an initial condition $\hat{x(0)} = 0$ and $P(0) = \text{covariance matrix of: } \hat{x(0)}$, KF provides a recursive calculation of state variables using minimized mean-squared error. The residuals between the measurement $z(t+1)$ and the estimation $H \hat{x}(t+1)$ is tested against predefined threshold $\tau^k$. If the residual exceeds the predetermined threshold, it is perceived as a bad data. The KF estimator generates better estimations of state variables than WLSE based estimator \cite{r25}; however conventional FDIA may impact the estimation performance, which can be avoided by updating measurement weighting function $W(t) = R^{-1}$ using the following equation \cite{r28} :

\begin{equation} \label{KF2}
    (W_{new}(t))^{-1} = (W(t))^{-1} \times e^{|z(t) - H \hat{x}(t|t-1)|}
\end{equation}
\\
The updated weighting function in eqn \ref{KF2} increases the FDIA detection probability for the deviation based KF approach compared to the conventional KF estimator \cite{r24}.

Hankel-matrix based bad data detection is another useful tool to detect the existence of any anomaly in time-series measurements, which looks for bad data over moving time-window \cite{rt}. Since the proposed attack scheme initiates a gradual incremental variation in phase angle over time, Hankel-matrix based BDD is an useful tool to detect such attacks. However, since the proposed attack model modifies the phase angle measurements very slowly over time, and the phase angle can deviate sharply during load change of peak-hour, it is not sufficient to observe low-rank approximation error of Hankel-matrix alone for the affected node to conclusively determine the occurrence of any attack.

\subsection{Proposed Detection Model}\label{detect}

In addition to the detection model proposed in \cite{rt}, we have exploited the spatial relation among the affected node and its topologically neighboring nodes. Phase angle measurements variations during GPS-spoofing attack can be identified using the gradient change in the low-rank approximation error of unwrapped phase angle measurements \cite{rt}. The key idea is: since the unwrapping process, by linearizing phase angle measurements, cause a monotonous decrease in low rank approximation error. Since any change in horizontal time-axis created by GPS-spoofing attack will cause a shift in the transition between $+/- \pi$ to $-/+ \pi$, the unwrapped phase angle measurement will demonstrate distortion in time series, subsequently causing a sharp change in gradient. However, the gradient change of low-rank approximation error of a single node is not enough evidence that the measurements from that node are under attacked.

For any load variation driven angle deviation of a particular node, it is expected that such angle-variation will be propagated to the neighboring location of the grid. Therefore, the affected node should show similar gradient change, either positive or negative, as the  neighbouring nodes in low-rank approximation error of unwrapped phase angle measurements. For GPS-spoofing attack targeting a particular node, the gradient change in low-rank approximation error for that particular node is opposite to the gradient change in low-rank approximation error of neighboring or topologically connected nodes. We can analytically express the relation between the change of gradient in low-rank approximation error among node $i$ and all other $\textit{M}$ nodes that are physically connected to $i$ as below:

\begin{align}
\frac{Gr_i}{|Gr_i|} \neq \frac{Gr_j}{|Gr_j|}\\
    Gr_k = \frac{\Delta L_k}{\Delta t}
\end{align} \\
Where $L_k$ = Unwrapped voltage/current phase angle measurements of node $k$.\\

\section{Experiment Method}\label{setup}

The proposed undetectable attack has been tested on  IEEE 24 bus reliability test system \cite{r32}. The physical power system is simulated in PSS/E.

The attack vector $a$ corresponds to the voltage phase angle. The attacker calculates the phase angle deviation required to instigate undetectable attack described in section \ref{formu}, and then creates the corresponding time shift by spoofing the GPS 1 PPS signal. The relation between the time shift $t_d$ and corresponding specific angle variation $\Delta \theta$ can be written as:

\begin{equation}\label{delay}
    \Delta \theta = 2 \pi f_0 t_d
\end{equation}

The optimization criterion described in eqn \ref{final} is solved using MATLAB $\textbf{Yalmip}$ tool box. For each time instance, the phase angle shift required to satisfy all the constraint in eqn \ref{final} is computed by using the $\textbf{Yalmip}$ solver. The attack vector $a$ represents the required phase angle shift $\Delta \theta$ to incur successful attack. The attack vector $a$ calculated with optimization equation is added to the phase angle and power measurements extracted from PSS/E. 
%The detectibility of the proposed attack model is tested using two robust residual based state estimation methods: Weighted Least Square (WLS) \cite{7}  and Deviation based Kalman Filtering (DKF) \cite{r24} method. The general Kalman Filtering approach \cite{r25} can also be a useful tool for state estimation, however the DKF method proposed in \cite{r24} provides more robust detection. So it is sufficient to perform the state estimation with DKF only, rather than with both of DKF and general KF. Additionally, $\chi^2$-test on the residual of WLS method is also performed.

The attacker solves the optimization algorithm of eqn. \ref{final} at each time instance using the phase angle measurements predicted $T_s$ sec. before corresponding time instance. Before analyzing the attack model, the accuracy of missing data recovery model to estimate future measurements needs to be verified. To check the accuracy of measurement estimation, voltage phase angle measurements are predicted consecutively. Hankel-matrix based missing data recovery model requires several predetermined parameters. Generally, longer measurement matrix is preferable, therefore a large value is preferable \cite{rec8} for the Hankel-matrix length $\tau$. We have tested a varying $\tau$ values to check the estimation accuracy. The number of rows $\kappa$ in Hankel-matrix depend on the matrix- length, a general rule of thumb being $\kappa = \tau/2$. Ref \cite{rec8} proposed eqn. \ref{hank_thr} to choose the estimation threshold of Hankel-matrix $\tau^e$, particularly for voltage phase angle.

\begin{equation}\label{hank_thr}
    \tau^e = max(2,30 e^{-3(t_i-t_d)}) \cdot \sum_{i=2}^L \frac{|\theta(i,t_i) - \theta(i,t_i-1)|}{L-1}
\end{equation}

where L refers to the number of initial trusted measurements, t is time instance, $t_d$ is the latest bad data occurrence time. The choice of L depends on the attacker. Since the measurement matrix is of length $\tau$, and the first estimation at time instance $\tau+1$ is performed with previous $\tau$ measurements; the number of initial trusted measurments can be equal to $\tau$.

To evaluate the undetectability of proposed phase angle shift caused by GPS-spoofing attack, we have utilized the aforementioned BDD methods: conventional WLSE residual test, Deviation based Kalman Filtering based residual test (DKF) and Hankel-matrix based estimation residual test. The residuals between the actual measurement and estimation using each method is tested with a predetermined threshold. The threshold is determined using Largest Normalized Residual test \cite{i2}. The predetermined threshold $\beta$ is chosen as 3 \cite{i2}. Finally, we test the effectiveness of the gradient-change in low-rank approximation error among neighbouring node-PMUs, as discussed in subsection IV-A.

For IEEE 24 bus RTS used for this work, load ($L_1$) at bus 13 is increased incrementally by 50$\%$ of $L_1$  at 5 sec., at 9 sec., and at 13 sec. In this way, a dynamic load varying condition is considered which behaves similarly as off-peak and peak-hour of real-world load variation. Here 13 sec. to 15 sec. duration is considered as peak-hour. The attacker instigate a small shift in GPS 1 PPS over time period T, such that the perceived power flow through the branch 13-23, which is calculated from DC power flow using the shifted phase angle measurements at each end of the branch , increases. The target is to make a significant impact on the calculated power flow during the peak load after 13 sec. At this time, the calculated power flow through the branch 13-23, perceived by SO, will cross the critical MVA limit of the branch, even though the actual MVA is within the limit. The MVA rating of the line 13-23 of IEEE 24 bus RTS is 500 MVA \cite{r32}.

Since the attacker instigates small incremental deviation in phase angle over time, it requires a long time for power flow to become large enough to cross MVA limit. As a result, attacker starts the attack at much earlier time of the peak-hour. In our simulation, the attacker starts the GSA from 2 sec.

\section{Result}

Before evaluating the proposed undetectable attack model, the Hankel-matrix based data estimation model needs to be verified. The attacker predicts future measurements to perform the optimization algorithm in order to create the proposed undetectable attack. For L number of initial trusted measurements, the attackers starts estimating the $(L+1)^{th}$ measurement. The estimation accuracy for continuous time instances is expected to decrease with time. Nevertheless, it is necessary to determine the number of continuous measurements that can be estimated accurately by the proposed model.

To test the feasibility of the data estimation model, we have performed continuous estimation, starting at $(L+1)^{th}$ time instance, of voltage phase angle measurement of bus 3 from the power system. A random Gaussian noise with mean 0 and variance 0.5 is added to the measurement to reflect the measurement noise in real world. 

\begin{figure}[hbt!]
 \includegraphics[height=2.1in,width=0.5\textwidth]{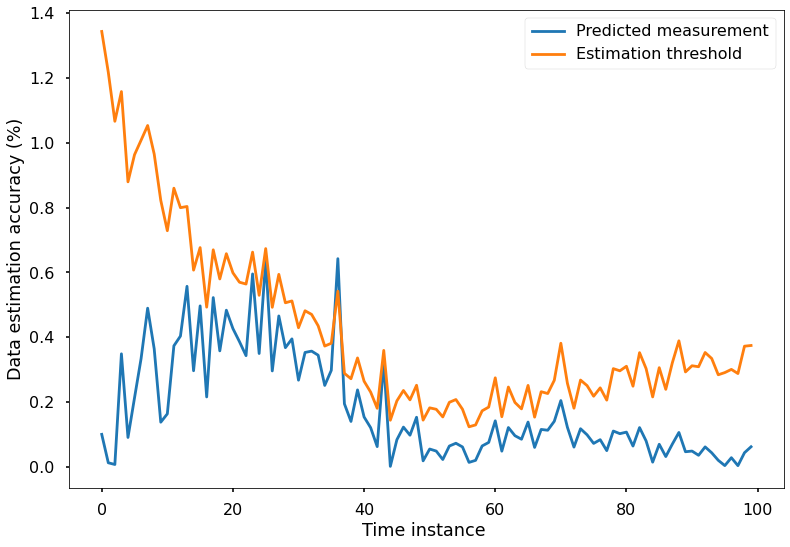}
 \caption{Estimation accuracy with initial trusted measurement number, L = 20}
 \label{es1}
\end{figure}

\begin{figure}[hbt!]
 \includegraphics[height=2.1in,width=0.5\textwidth]{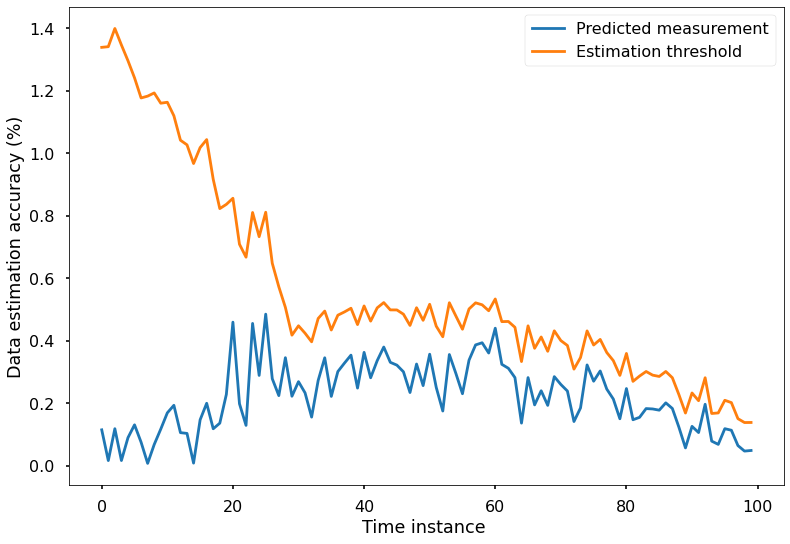}
 \caption{Estimation accuracy with initial trusted measurement number, L = 30}
 \label{es2}
\end{figure}

\begin{figure}[hbt!]
 \includegraphics[height=2.1in,width=0.5\textwidth]{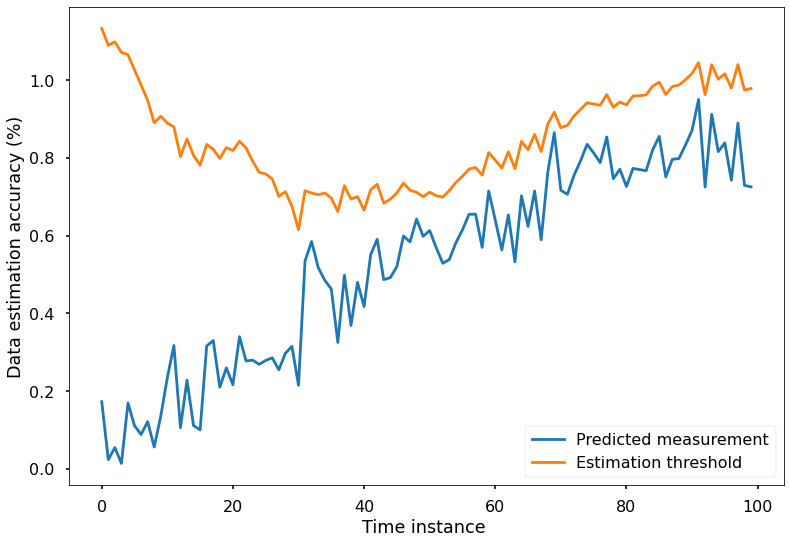}
 \caption{Estimation accuracy with initial trusted measurement number, L = 50}
 \label{es3}
\end{figure}

All the data in the trusted measurement window after $2L$ time instances are estimated measurements at previous time instances, As a result it is expected that the gap between estimation accuracy and the threshold will be smaller. This is confirmed from the results illustrated in fig \ref{es1} to \ref{es3}. The optimization algorithm of eqn \ref{final} requires approximately $T_s = 0.15$ sec. of computational time. Since the PMU transmit 60 measurements per sec., it is sufficient for the attacker to estimate $0.15 \times 60 = 9$ measurements accurately, to ensure the formulation of proposed attack model at time instance $t_i$. From the figures \ref{es1} to \ref{es3}, the estimation model can accurately estimate measurements with substantial gap between the accuracy and threshold upto 25 continuous time instance for the case of $L=50$. The attacker implements the estimation model with 50 initial trusted measurements.

\subsection{Impact on Power Flow measurement}

At first, starting at 2 sec., the proposed optimization algorithm for each time instance is executed and the corresponding optimal attack vector $a_{op} = \theta_0$, which is the phase angle shift required to achieve the target of misleading the power flow calculation, is determined. In the rest of the paper, the term attack vector and phase angle shift will be used interchangeably.  Time-shifts (as in eqn. \ref{delay}) are applied to the GPS 1 PPS signal of the corresponding PMUs located at the bus 23, simulated by adding corresponding optimized attack value $a_{op}$ to the voltage phase angle measurements of bus 23. The power flow through the branch connecting the buses 13 and 23 is calculated using DC power flow equation. Since there is a presence of time shift in the GPS 1 PPS into the corresponding PMUs at the both ends of the branch connecting buses 13 and 23, the perceived power flow through the branch will be different from the actual value.

In the second case, two random time shifts are applied to GPS signals received by the corresponding PMUs located at buses 3 and 18 and the corresponding power flow is calculated at each time instance, starting from 5 sec. The attack vector in this case is $a_{rand}$.

\begin{figure}[hbt!]
 \includegraphics[width=0.49\textwidth]{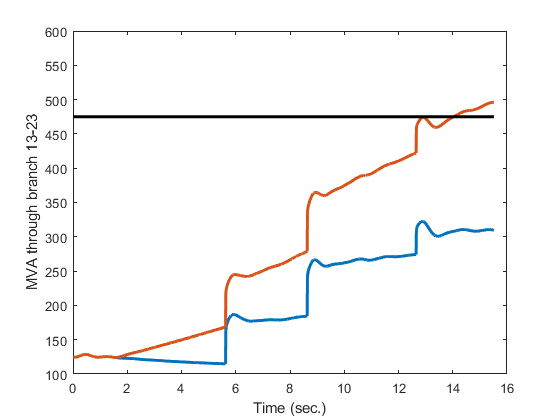}
 \caption{Power flow calculation between bus 13 and bus 23 for $a_{op}$ over T time period}
 \label{pf1}
\end{figure}

Fig. \ref{pf1} demonstrates the results obtained from applying GPS-spoofing attack using proposed attack model. From fig. \ref{pf1}, we can observe that the perceived power flow calculated using the shifted phase angle measurement is higher than the actual power flow through the branch 13-23. for each time instance after initiating attack at 2 sec. When the power flow through the branch increases, the calculated power flow is also increased by a small amount, keeping the calculated power flow within the line flow limit. After 14 sec, when the maximum power flows through the branch at peak hour, the calculated MVA exceeds the critical line flow limit, which is $95\%$ of 500 MVA. Therefore the control center, thinking that critical MVA limit has been exceeded, will be forced to take contingency actions despite the actual flow is within the limit, and the branch 13-23 will be tripped.

From fig. \ref{pf1}, we can conclude that the attacker can successfully cause change gradual change in power flow calculation by initiating an incremental GPS-spoofing attack over a long time, and the attacker can make the perceived flow calculation by SO exceed the flow limit for a particular branch, forcing the SO react against a false alarm of critical flow limit breach.

\subsection{Undetectibility Analysis}

\begin{figure}[hbt!]
 \includegraphics[width=0.5\textwidth]{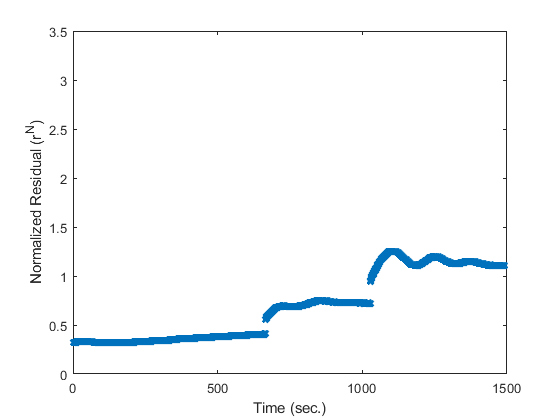}
 \caption{Normalized WLSE residuals between the observed and expected current measurements through the branch between bus 13 and 23}
 \label{kf1}
\end{figure}

\begin{figure}[hbt!]
 \includegraphics[width=0.5\textwidth]{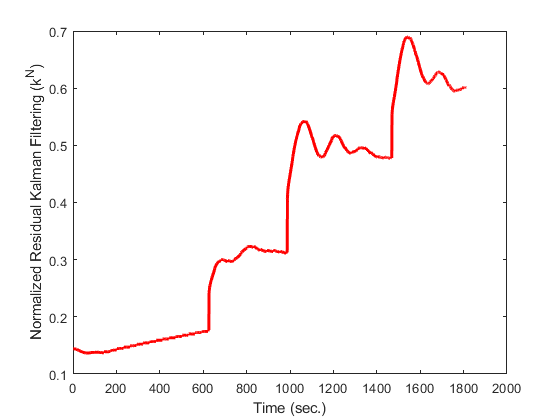}
 \caption{Normalized DKF residuals between the observed and expected current measurements through the branch between bus 13 and 23}
 \label{rs1}
 \end{figure}

To check the detectibility of proposed attack scheme, normalized estimated residuals at each timestamp is calculated using WLSE and DKF methods separately. The state variable, $X$, is a $24 \times 1$ matrix comprised of voltage phase angles, and the Jacobian matrix, $H$, is $38 \times 24$ bus admittance matrix. The current measurement variable, $Z$, is a $38 \times 1$ matrix representing the current flow through the 38 branches of IEEE 24 bus RTS.   %Fig \ref{wlsi} and \ref{wlsr} demonstrate the results obtained from WLS bad data detection technique on the imaginary and real components, respectively.  Both figures show that, the estimation residuals are mostly less than estimation threshold, particularly when the threshold is $\mu + 3 \sigma$. 
Fig. \ref{rs1} and \ref{kf1} demonstrate the normalized estimation residuals obtained using WLSE and DKF methods, respectively. The normalized estimation residuals $||r||_2\||\Omega||_2$ at each timestamp is observed. From the results, it is evident that the largest normalized residual is less than threshold $ \beta = 3$ for both WLSE and DKF methods. 


Results from fig. \ref{rs1} and \ref{kf1} indicates the undetectibility of the proposed attack model using two of most commonly used robust detection techniques: WLSE and DKF.

Hankel-matrix based detection model proposed in \ref{rt} detects distortion in unwrapped phase angle measurements to determine the possiblity of GPS-spoofing attack. To analyze the effectiveness of Hankel-matrix model, we have extracted unwrapped phase angle measurements for both of normal condition, period of time, and of gradual GPS-spoofing attack condition discussed in this article over same 15 sec. time period. Estimation error using Low rank approximation with moving time window of length $W = 80$, $W = 100$ and $W = 120$.

\begin{figure}[hbt!]
 \includegraphics[width=0.5\textwidth]{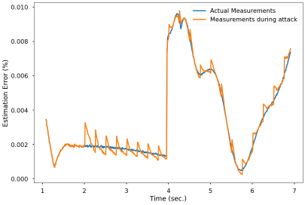}
 \caption{Estimation error of Hankel-matrix under normal and attack conditions over moving time window, W = 80}
 \label{hank1}
 \end{figure}
\begin{figure}[hbt!]
 \includegraphics[width=0.5\textwidth]{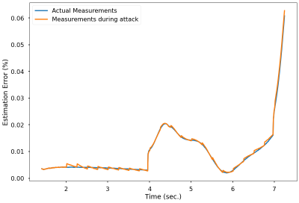}
 \caption{Estimation error of Hankel-matrix under normal and attack conditions over moving time window, W = 100}
 \label{hank1}
 \end{figure}
 \begin{figure}[hbt!]
 \includegraphics[width=0.5\textwidth]{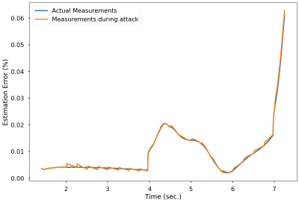}
 \caption{Estimation error of Hankel-matrix under normal and attack conditions over moving time window, W =120}
 \label{hank1}
 \end{figure}

All three cases of different time-window lengths W, the attack condition demonstrates spikes in estimation error using low-rank approximation of Hankel-matrix. However, similar spike can be observed for phase angle deviations due to load changes too. Therefore, the spikes in fig. \ref{hank1} to \ref{hank3} are not conclusive evidence of GPS-spoofing attack.

\subsection{Effectiveness of Gradient of Hankel-Matrix Model}

As discussed in subsection \ref{detect}, the comparison of change in gradient of low-rank approximation error among neighboring PMUs can indicate small deviation in phase angle measurements of a particular PMU measurements, implying a possible GPS-spoofing attack.

 \begin{figure}[hbt!]
 \includegraphics[width=0.5\textwidth]{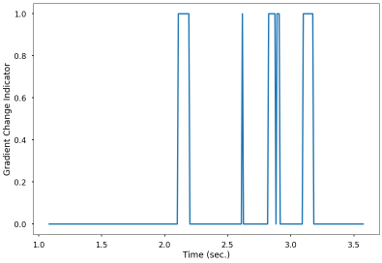}
 \caption{Change in gradient of low-rank approximation error between Bus 13 and 23}
 \label{grd1}
 \end{figure}
\begin{figure}[hbt!]
 \includegraphics[width=0.5\textwidth]{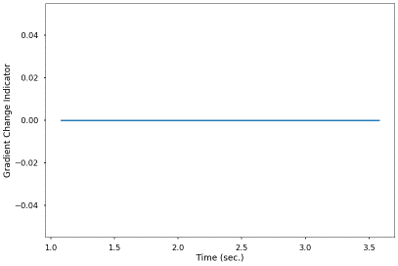}
 \caption{Change in gradient of low-rank approximation error between Bus 13 and 12}
 \label{grd2}
 \end{figure}

The change in gradient of low-rank approximation error between two buses is demonstrated with boolen expression. The value $1$ signifies a change in gradient of low-rank approximation error between two buses, and the value $0$ signifies unchanged condition. Fig. \ref{grd1} shows that, the gradient changes frequently after 2 sec. among the low-rank approximation errors of Hankel-matrices of bus 13 and 23, using unwrapped phase angle measurements. However, gradients of low-rank approximation errors of Hankel-matrices using unwrapped phase angle measurements does not demonstrate any change among bus 13 and 12 (fig. \ref{grd2}). Similar changes are observed among bus 23 and 12, and bus 23 and 11, separately. The results imply the effectiveness of gradient of Hankel-matrix model in the detection of proposed gradual undetectable GPS-spoofing attack.
 
\section{Conclusion}
Despite providing efficient and accurate time-synchronized data, the Phasor Measurement Unit (PMU) is prone to various types of cyberattacks. One of the most critical cyber attacks against PMU is the GPS-spoofing attack, where the reference signal for time-tagging is modified. This paper aims at developing a stealthy GPS-spoofing attack, where the attacker inject very small scale shift in the GPS 1 PPS signal at each timestamp, and gradually increases the deviation. Since the effect of timestamp shift is reflected by a shift in phase angle, the perceived power flow through a branch calculated using the phase angle at the both ends of the branch is changed, even though the actual power flow through the same branch remains unchanged. The attacker's goal is to increase the perceived power flow, and make the perceived power flow exceeds the line flow limit at the peak hour.

At first, the paper proposes an undetectable relentless small incremental GPS-spoofing attack scheme which is able to bypass the Bad Data Detection (BDD) algorithm. The attack is applied by solving a convex optimization criterion at regular time interval, so that after a specific time period the attack vector incurs a significant change in the perceived power flow calculated with shifted phase angles, hence forcing the SO take contingency actions. Secondly, the proposed attack model is tested on the IEEE 24 bus RTS in PSS/E. The corresponding impact on the calculated power flow through the branch connecting buses 13 and 23 is observed. Load at bus 13 is varied to reflect the off-peak and peak energy demand. Thirdly, the undetectability of the proposed algorithm is tested against WLS, DKF and Hankel-matrix. The result shows that the attacker can modify the perceived power flow through the branch 13-23 and make it cross the line flow limit at the peak hour (13 sec.). Moreover, it can be observed that the proposed attack remains undetected against WLSE and DKF methods. Even though Hankel-matrix model with unwrapped phase angle measurements demonstrates spikes during gradual deviation of phase angle over time, it cannot detect the proposed attack conclusively. Proposed novel detection model, that compares the gradient of low-rank approximation of Hankel-matrix among topologically connected node measurements, can conclusively detect relentless small incremental GPS-spoofing attack.

\bibliography{ref1} 
\bibliographystyle{ieeetr}
\end{document}